\documentstyle[12pt]{article}
\newcommand{\be}{\begin{equation}}
\newcommand{\ee}{\end{equation}}
\def\n{\noindent}

\catcode `\@=11
\@addtoreset{equation}{section}
 
\catcode `\@=12

\title{\bf\huge On ``minimally curved spacetimes'' \\
in general relativity}     
\author{Naresh Dadhich\thanks{E-mail : nkd@iucaa.ernet.in} \\
{\sl Inter-University Centre for Astronomy \& Astrophysics,}\\
{\sl Post Bag 4, Ganeshkhind, Pune - 411 007, India.} 
} 
\date{}
\begin{document}
\maketitle

\begin{abstract}

\n We consider a spacetime corresponding to uniform relativistic
potential analogus to Newtonian potential
as an example of ``minimally curved spacetime''. We also consider a
radially symmetric analogue of the Rindler spacetime of uniform proper
acceleration relative to infinity.
\end{abstract}


\vfill
\begin{flushright} IUCAA-37/97 \end{flushright}
\vfill

\newpage

\n It is rather a difficult question to characterise  minimality  of 
curvature  for  spacetime, because there is  no  clear  geometric 
criterion.  The  only  invariant  geometric  criterion  in   this 
connection is of constant curvature which specifies the de Sitter 
spacetime.  Let  us  begin  with  some  physical   considerations 
involving  the standard canonical setting of radial symmetry.  In 
the  Newtonian theory (NT) radially symmetric field of a body  at 
rest  is specified fully by the potential, $\phi = k - M/r$,  where   
$M$ refers to mass of the body and $k$   provides freedom to set zero of 
the potential wherever one wishes. What could be the  minimality 
condition   for  this  field  ?  The  obvious  answer  would   be            
$\phi = k = const.$, which really means absence of the field.
And hence in NT  there cannot be any condition defining 
minimality of radially symmetric 
field. \\

\n The question is can there be one in general relativity (GR) ? The 
question is worth following for in GR gravitation is described by 
curvature  of  spacetime which is given by 20 components  of  the 
Riemann curvature tensor. This obviously is a larger and  richer 
system  which should in principle have greater  flexibility  than 
NT. First thing to consider is what happens when analogue of  the 
Newtonian   potential  is  set  to  constant  in  the   spacetime 
describing  gravitational field of a massive body at rest ?  Does 
it lead to flat spacetime indicating absence of gravity as in  NT 
?  If it does not, then it may qualify for minimality.  It  turns 
out  that  it  does  not [1-2] and hence  we  propose  to  define 
minimally curved spacetime corresponding  to a constant but  non-
zero  relativistic  potential, by which we shall  mean  a  scalar 
function  that completely determines the field and goes  over  to 
the Newtonian potential in the limit. \\
 
\n Gravitational acceleration can be  countered 
locally  by appropriate choice of a frame. It cannot  however  be 
anulled  globally. The Rindler spacetime [3]  represents  uniform 
acceleration  in a particular direction, which can be removed  by 
a  proper  choice  of a frame and hence its spacetime is flat.  How  about 
considering   a   spacetime  that   represents   uniform   radial 
acceleration  ? It would obviously be  non-flat  because  though 
acceleration  is uniform but it is radial and hence in  no  frame 
can  it vanish.  This would be a candidate for one  rung lower in 
the order of minimality of curvature. It would correspond to  a 
homogeneous  radial  field in the Newtonian limit. Thus  we  have 
minimally curved spacetime that corresponds to absence of gravity 
in the Newtonian limit. The next in the line would be a spacetime 
corresponding to uniform radial field in the Newtonian limit. \\
 
\n It  is however remarkable to note that both these  spacetimes  in 
the  curvature coordinates have all but one curvature  components 
zero.  This  is how the author has first found  them long ago 
[4], following a tentative enquiry of finding spacetimes with the 
least  number  of curvatures non-zero.  This is not  a  covariant 
criterion.  The amazing thing is that spacetimes so obtained   do 
seem  to accord to what was roughly asked of them. The  covariant 
properties characterising the spacetimes are  vanishing of proper 
acceleration (indicating zero active gravitational mass) for  the 
former  while  uniform proper acceleration relative  to  infinity 
for the latter. \\
    
\n Let us begin with a spherically symmetric metric, 

\be
ds^2 = Bdt^2 - A dr^2 - r^2 d \Omega^2, d \Omega^2 = d \theta^2 +
sin^2 \theta d \varphi^2
\ee

\n where  $B$ and $A$ are functions of $r$   and $t$  in  general. 
For  empty space let us consider $R_{01} = 0$ and $R^0_0 = R^1_1$, 
which will lead to $B = A^{-1} = 1 + 2 \phi (r)$  and 
consequently we can write [1,2]

\be
R^0_0 = - \bigtriangledown^2 \phi = - \frac{1}{r} 
(r \phi)^{\prime \prime},
\ee

\be
R^2_2 = - \frac{1}{r^2} (r \phi)^{\prime}, \phi^{\prime} = d \phi/dr.
\ee

\n Note  that $R^2_2 = 0$  is  the  first  integral  of $R^0_0 = 0$, 
the Laplace 
equation.  This  has the general solution $\phi = k - M/r $. It  is 
$R^0_0 = 0$ that  determines  the free parameter $k = 0$  and  the  Schwarzschild 
solution follows. It is this function $\phi$  that determines the field 
entirely and is the analogue of the Newtonian potential. We shall 
term  it  relativistic  potential. That  means  the  analogue  of 
constant  potential  in NT would be $\phi = const. = k$. As  is 
clear from (3) that it will give rise to a non-flat spacetime, 

\be
ds^2 = dt^2 - (1 + 2k)^{-1} dr^2 - r^2 d \Omega^2
\ee

\n where  the  factor $(1 + 2k)$  has been absorbed by 
redefining $t$.  This 
gives rise to 

\be
R^{23}_{~~23} = \frac{2k}{r^2} = -R^2_2,~ T^0_0 = T^1_1 = - \frac{k}
{4 \pi r^2} .
\ee

\n The metric (4) hence qualifies for minimally curved spacetime (MCS)
for it  is 
free  of  acceleration as well as tidal acceleration  for  radial 
motion and it has zero gravitational mass indicated by $\rho_c = T^0_0
- T^{\alpha}_{\alpha} = 0$. It 
may  be  noted  that  weak field but  no  restriction  on  motion 
(special    relativistic)   limit   of   GR   should   read    as                                        $R^0_0 = -\bigtriangledown^2 \phi = -4 \pi \rho_c (\rho_c = \rho + 3p$
for perfect fluid). It is vanishing of $\rho_c$ implies zero gravitational
charge. The sole
surviving curvature is $R^{23}_{~~23} = 2k/r^2$, 
which will make its presence felt 
in tidal acceleration for transverse motion only. This  curvature 
is an invariant for spherical symmetry; i.e if it alone is  non-
zero  in  one  coordinate  system  then it  will  be  so  in  all 
coordinates preserving spherical symmetry. \\

\n In  GR relativistic potential $\phi$  not  only  imparts 
acceleration  to free particles but it also curves space to  take 
into account non-linear aspect of the theory [2]. In the process 
$\phi$ exlicitly  occurs  in  the Riemann  curvature  tensor  lending 
physical meaning to itself. It is this that allows us to have a  non-
flat   spacetime   corresponding   to   constant   but   non-zero 
relativistic  potential $\phi$,  which we wish to  designate as  minimally 
curved.  The absolute zero of relativistic potential  is  however 
defined  by  flat  spacetime.  It is  interesting  to  find  that 
stresses 
in  (5)  are  exactly what are required  to  represent  a  global 
monopole  [5].  A global monopole is supposed to  result  from  a 
spontaneous  breaking of global $O(3)$ symmetry into $U(1)$ in  phase 
transition in the early Universe. Thus MCS (4) 
finds application in a very exotic setting. \\

\n There  is  a  novel way to construct the  metric  (4)  through  a 
geometric ansatz. Consider a 5-Minkowski spacetime 

\be
ds^2 = dt^2 - dx^2 - dy^2 - dz^2 - dw^2
\ee

\n and then impose the restriction 

\be
x^2 + y^2 + z^2 + w^2 = \alpha^2 ( x^2 + y^2 + z^2)
\ee

\n where $\alpha$  is a constant. This generates a curved spacetime which is 
nothing but MCS (4). This ansatz could 
be  extended by letting other variables participate and it is  in 
fact a prescription to generate spacetimes of zero  gravitational 
mass $(T^0_0 - T^{\alpha}_{\alpha} = \rho + 3p = 0$) [5]. \\

\n Let  us  now come from zero acceleration to the case  of  uniform 
acceleration. It is clear that acceleration is determined by  the 
metric potential $B(r)$ alone and hence we set $A = 1$ in the  metric 
(1)  so  that  the space part is flat.  The  proper  acceleration 
relative to infinity is given by $B^{\prime} B^{-1/2} = const. = 2a$.
This gives $B = (a r + b)^2$ where $b$ is a constant.   
We have thus spacetime of uniform acceleration given  by  the 
metric, 

\be
ds^2 = (a r + b)^2 dt^2 - dr^2 - r^2 d \Omega^2.
\ee

\n Note that this has the sole surviving curvature $R^{02}_{~~02}
= (a/r) (ar + b)^{-1}$ [4].  It 
is  clear  that uniform acceleration $a$ here
 cannot be transformed away as in  the  Rindler 
case [3]. It has stresses given by 

\be
T^1_1 = - \frac{a}{2 \pi r (ar + b)} = 2 T^2_2
\ee

\n all   others  including  energy  density  being   zero.   The 
gravitational charge density responsible for the acceleration  is              
$\pi \rho_c = (a/r) (ar + b)^{-1}$.
This  is  an  example  of  spacetime  that  incorporates  uniform 
acceleration.  It may have application in a situation  where  the 
source is undetectable except for its influence  on  matter - a 
true dark  matter.  There  have  been 
considerations  invoking  modification of the  Newtonian  law  of 
gravity [6] to explain the flat rotation curves of galaxies. This 
could perhaps be relativistic version of such considerations. The 
advantage  here is that source is in principle undetectable.  The 
hard question is how to justify such a consideration on physical 
grounds. It though offers an interesting possibility to speculate 
on. \\

\n We  have very recently considered an interesting  application  of 
MCS  (4)  in  a classical situation [7-8]. It has  been  used  to 
characterise isothermal $(\rho \sim 1/r^2$ and linear equation
of state) perfect fluid unbounded spheres. The necessary and 
sufficient  condition for a spherically symmetric unbounded perfect  fluid 
model to be isothermal is that the spacetime is conformal to  the 
MCS metric (4). In particular if we superpose spacetimes (4) and (8)
by writing $B = (ar + b)^2$ and $A = (1 + 2k)^{-1} $ in the metric (1),
we generate stiff fluid spacetime with $\rho = p = 1/16 \pi r^2$ for
$k = -1/4$ and $b = 0$. \\  

\n The other probing question it leads to is that no region in space
can be shielded from a constant potential which may be produced by the
matter-energy distribution in the rest of the Universe (ROU). That
is so long as ROU is non-empty, a local region will have an
underlying constant potential, which will not make any difference
in NT because it is physically inert. But its analogue in GR, relativistic
potential is not physically inert. Thus a local region will have underlying MCS (4)
as its asymptotic limit. ROU exterior to cavity can be taken in the large to
the homogeneous and isotropic which will produce uniform Newtonian
potential inside the cavity. This will give rise to an MCS (4) corresponding
to uniform relativistic potential. The asymptotic limit of the spacetime
inside the cavity will then be MCS (4) and not flat spacetime. This
is how MCS will relate  local regions with ROU. This is perhaps
the simplest way to be in consonance with Mach's Principle in its very
basic form, relating local to global. \\

\n This would however raise very deep and basic questions. In particular
asymptotic flatness is tied to Ricci flatness, giving up of which
would mean giving up of vacuum as well [2]. A local region can only relate
to ROU through its asymptotic limiting spacetime, which has to be non-flat
for a non-vacuous consideration. The prime question is how to
achieve consistency between the vacuum equations and asymptotic
non-flat character. We shall take up discussion of such questions elsewhere.

\end{document}